\documentclass[twocolumn,preprintnumbers,nofootinbib,prl,aps,10pt]{revtex4-1}
\usepackage{psfrag,graphicx,epsfig,amsmath,amssymb,bm}
\usepackage[dvips]{color}
\usepackage[caption=false]{subfig}
\usepackage{multirow}
\usepackage{natbib}
\usepackage{hyperref}
\def\beq{\begin{equation}}
\def\eeq{\end{equation}}
\def\beqa{\begin{eqnarray}}
\def\eeqa{\end{eqnarray}}

\def \pt {p_T} 

\def \psl{p{\!\!\!/}}
\def \ptm{p{\!\!\!/}_T}
\def \ttbar{t\bar{t}}

\begin{document}

\title{
Boosting Higgs CP properties via VH Production at the Large Hadron Collider
}

\author{Rohini Godbole\,$^a$, David J. Miller\,$^b$, 
Kirtimaan Mohan\,$^{a,c}$, and Chris D. White\,$^b$} 

\affiliation{$^a$ Center for High Energy Physics, Indian Institute of
  Science, Bangalore 560 012, India\\ $^b$ SUPA, School of Physics and
  Astronomy, University of Glasgow, Glasgow, G12 8QQ, UK\\ $^c$
  LAPTh. Univ.~de Savoie. CNRS. B.P.110, Annecy-le-Vieux, F-74941,
  France}
  
\preprint{LAPTH-031/13}

\begin{abstract}
\noindent
We consider $ZH$ and $WH$ production at the Large Hadron Collider,
where the Higgs decays to a $b\bar{b}$ pair.  We use jet substructure
techniques to reconstruct the Higgs boson and construct angular
observables involving leptonic decay products of the vector bosons.
These efficiently discriminate between the tensor structure of the
$HVV$ vertex expected in the Standard Model and that arising from
possible new physics, as quantified by higher dimensional operators.
This can then be used to examine the CP nature of the Higgs as well as
CP mixing effects in the $HZZ$ and $HWW$ vertices separately.
\end{abstract}


\maketitle

\noindent
The recently discovered Higgs-like
particle~\cite{Chatrchyan:2012ufa,*Aad:2012tfa} at the Large Hadron
Collider (LHC) will dominate global particle physics effort for many
years to come. Whether or not this is the Standard Model (SM) Higgs
boson necessitates precise study of its couplings to other SM
particles in all production and decay channels. In this letter we
focus on the determination of the tensor structure and hence the CP
properties of the vertex responsible for the production of a Higgs
boson associated with a $V=W,Z$ boson as well as CP violating effects
in the same.  We show that this process allows one to disentangle CP
even and CP odd new physics from the SM contribution (and from each
other).  Furthermore, one may probe the $HWW$ and $HZZ$ vertices {\it
  separately}.  The $VH$ channel, though subdominant, has been shown
to be viable due to the use of modern jet substructure
techniques~\cite{Butterworth:2008iy}. We show that, interestingly, the
very kinematic cuts that are required to make the detection of the
$VH$ channel viable at the LHC using this technique, automatically add
to its discriminatory power.

Corrections to the Standard Model $HVV$ vertex can be written by
supplementing the SM Lagrangian with higher dimensional operators that
can originate from Beyond the SM (BSM) physics:
\begin{equation}
g_W^2
\frac{c_1}{2\Lambda_1^2}\Phi^\dagger\,\Phi\,F_{\mu\nu}\,F^{\mu\nu}, \quad
g_W^2\frac{c_2}{2\Lambda_2^2}\Phi^\dagger\,\Phi\,\tilde{F}_{\mu\nu}\,F^{\mu\nu},
\label{Lagrangian}
\end{equation}
with $g_W$ and $F_{\mu\nu}$ the electroweak coupling constant and
SU(2) field strength tensor,
$\tilde{F}_{\mu\nu}=\epsilon_{\mu\nu\alpha\beta}F^{\alpha\beta}$,
$\Phi$ the (SU(2) doublet) Higgs field and ${c_i}$ (complex)
constants.  These operators (CP even / odd respectively) arise from
integrating out higher energy dynamics, are suppressed by mass scales
$\Lambda_i^2$, and modify the $HWW$ vertex to
\begin{equation}
ig_W M_W \left[g^{\mu\nu} 
+ \frac{4 c_1}{\Lambda_1^2}\left(p^{\mu}q^{\nu} -g^{\mu\nu}p \cdot q  \right)
+ \frac{8 c_2}{\Lambda_2^2}\epsilon^{\mu\nu\rho\sigma}p_{\rho}q_{\sigma}
 \right]
\end{equation} 
where $M_W$ is the $W$ boson mass and $p$ and $q$ the $W$ boson momenta. 

The nature of the $HZZ$ interaction in the four lepton channel was
investigated in
\cite{Choi:2002jk,*Godbole:2007cn,Gao:2010qx,DeRujula:2010ys,Bolognesi:2012mm,Stolarski:2012ps}
and has now been probed with current LHC data,
disfavouring a pure pseudoscalar hypothesis at \mbox{$\sim$
  2--3$\,\sigma$~\cite{ATLAS-CONF-2013-013,*CMS-PAS-HIG-13-002,Chatrchyan:2012jja}
}. Similar constraints on the $HWW$ vertex using the $H \to WW$ decay
are hard to achieve since the kinematical cuts necessary to eliminate
backgrounds hamper the analysis.  Although the tensor structure of the
$HVV$ vertex can be investigated using kinematic and angular
correlations~\cite{Plehn:2001nj,Zeppen:2006} in Vector Boson Fusion
(VBF), one cannot separately study the $Z$ and $W$ contributions in
this channel.  Furthermore, non-SM $HVV$ vertices have a reduced
acceptance to VBF-like kinematic cuts.  Electron-positron colliders
with polarized beams can offer precision information on the $HZZ$
vertex (e.g. CP structure), via angular distributions of leptonic
decay products of the vector
bosons~\cite{Miller:2001bi,*Han:2000mi,*Rindani:2009pb,Biswal:2008tg,*Biswal:2009ar,*Dutta:2008bh}.
However, the determination of the $HWW$ vertex
\cite{Biswal:2008tg,*Biswal:2009ar,*Dutta:2008bh}, possible only by
studying $e^{+}e^{-} \rightarrow \nu \bar \nu H$, has an irreducible
background from $ZH$ production followed by $Z \rightarrow \nu \bar
\nu$.  Thus an unambiguous separate determination of the $HWW$ and
$HZZ$ vertices through VBF is possible only at the proposed
LHeC~\cite{Biswal:2012mp}. In order to elucidate the nature of the
$HVV$ couplings at the LHC, one is unavoidably led to $VH$ production.

At the LHC, where until recently even the detection of the Higgs in
the $VH$ channel was considered difficult, studies of the nature of
the $HVV$ vertex were not contemplated.  Here we show that modern jet
substructure techniques~\cite{Butterworth:2008iy}, which probe the
kinematical region where the Higgs is highly boosted and decays to a
$b\bar{b}$ pair, increase the sensitivity to BSM couplings.  We
furthermore demonstrate that angular correlations of decay leptons
produced in the $VH$ process, are able to distinguish between the
different contributions in eq.~(\ref{Lagrangian}).  We simulate all
processes using {\tt MadGraph5}~\cite{madgraph} interfaced with {\tt
  Pythia6}~\cite{pythia} and use the {\tt FastJet}
package~\cite{fastjet} to cluster the jets. The effective Lagrangian
was implemented in {\tt FeynRules}~\cite{feynrules}.

\section{Event Selection}

It is important to apply selection criteria to distinguish between the
signal and background processes. For $ZH$ production we require:
\begin{enumerate}
\item A fat jet (radius $R=1.2$, $p_T>200\,{\rm GeV}$). After applying
  the filtering procedure of~\cite{Butterworth:2008iy}, we require no
  more than three subjets with $p_T>20\,{\rm GeV}$,
  \mbox{$|\eta|<2.5$}, and radius $R_{sub}=\min(0.3,R_{bb})$, 
  where $R_{bb}$ is the separation of the two hardest subjets, both of
  which must be $b$-tagged. 
\item Exactly 2 leptons (transverse momentum $\pt>20\,{\rm GeV}$, 
pseudo-rapidity $|\eta| <2.5 $) of same
  flavour and opposite charge, with invariant mass within
  $10\,{\rm GeV}$ of the $Z$ mass $M_Z$. These should
  be isolated: the sum of all particle transverse
  momenta in a cone of radius $R=0.3$ about each lepton should not
  exceed 10\% of that of the lepton.
\item Demand that the reconstructed $Z$ has a $\pt >150\,{\rm GeV}$,
  with azimuthal angle satisfying $\Delta\phi(Z,H)> 1.2$.
\end{enumerate}
After cuts, the only significant surviving background process is $Z$ +
jets.  Cross-sections at Leading Order (LO) before and after the cuts
are shown in Table~\ref{tab:cs1}. The $H\to b\bar{b}$ branching ratios
were taken from Ref.~\cite{LHCH:2011ti}. Since the $K$ factors for the
background and the signal are not too
different~\cite{Butterworth:2008iy} our results are not expected to be
significantly different at Next to Leading Order (NLO).  In principle
our analysis is sensitive to the $b$-tagging efficiency and light
quark jet rejection rate~\cite{Butterworth:2008iy} (here set to
$\epsilon_b=0.6$ and $r_{j}=100$ respectively). However, we checked
that this has no impact on the angular observables.

\begin{table}
\begin{tabular}{|c|c|c|c|c|c|c|c|}\hline
Channel & $VH_{SM}$ & V+jets & $\ttbar$ & Single top &
$VH^{0^{+}}_{BSM}$ & $VH^{0^{-}}_{BSM}$ \\ \hline
ZH & 0.153 & 0.416 & 0    & 0    & 0.61 & 0.93 \\\hline
WH & 0.455 & 0.33  & 0.16 & 0.06 & 1.86 & 2.74 \\ \hline                             
\end{tabular}
\caption{Cross-sections (fb) evaluated at leading order
for the $14\,{\rm Te}$V LHC after applying all cuts. $V$+jets 
corresponds to the $Z$+jets background for the $ZH$ process 
and $W$+jets for the $WH$ process. For the last two columns the 
SM contribution was set to zero and the values of $\Lambda_1$ 
and $\Lambda_2$ were set to 
reproduce the SM total cross-section before applying cuts. 
These results do not require $W$ reconstruction.}
\label{tab:cs1}
\end{table}

For $WH$ production we require:
\begin{enumerate}
\item The Higgs reconstructed as above.
\item Exactly 1 hard lepton ($\pt > 30\,{\rm GeV}$, $|\eta| <2.5 $),
  isolated as above.
\item Missing transverse momentum $\ptm > 30\,{\rm GeV}$.
\item The reconstructed $W$ has $\pt > 150\,{\rm GeV}$ and azimuthal
  angle satisfying $\Delta\phi(H,W) > 1.2$.
\item No additional jet activity with $\pt^{jet}>30\,{\rm GeV}$,
  $|\eta|< 3$ (to suppress single and top pair production
  backgrounds).
\end{enumerate}
Again, major backgrounds are detailed in Table~\ref{tab:cs1}.

\subsection{\label{sec:w-reco}Reconstructing the neutrino momentum}

One must reconstruct the neutrino in $WH$ production to
determine our angular observables.  We identify
the neutrino transverse momentum $\vec{p}_{T\, \nu}$ with the missing
transverse momentum $\vec{\psl}_T$, and demand the squared sum of the neutrino
and lepton momenta be equal to the squared $W$ boson
mass $((\vec{p}_{\nu}+\vec{p}_{l_1})^2=M_{W}^2)$, solving the
resulting quadratic equation.  Comparing with the ``true''
Monte-Carlo generated neutrino momentum, we find that choosing 
a given solution out of the two possible ones,
reconstructs the true neutrino
momentum $50\%$ of the time, with $\simeq5\%$ giving 
imaginary solutions.  One can also compare the boosts
of the Higgs $\beta_{z}^{H}$ and reconstructed $W$ $\beta_{z}^{W}$
in the $z$ direction. The solution with the minimum value for
$|\beta_{z}^{W} - \beta_{z}^{H}|$ 
gives the true neutrino momentum in $65\%$ of cases.

We present results for angular observables for the three
cases listed below and show that the differences are small:
\begin{itemize}
\item Choose the solution by demanding the difference in boost of $W$
  and $H$ is minimized (BT).
\item Use the ``Monte-Carlo Truth'' (MCT), where the neutrino momentum
  is reconstructed from the $\pt$ of the ``true'' neutrino rather
  than the $\ptm$.
\item Use both solutions of the quadratic equation (BN).
\end{itemize}

\subsection{Sensitivity to new operators}

\begin{figure}
\centering
\includegraphics[width=0.4\textwidth]{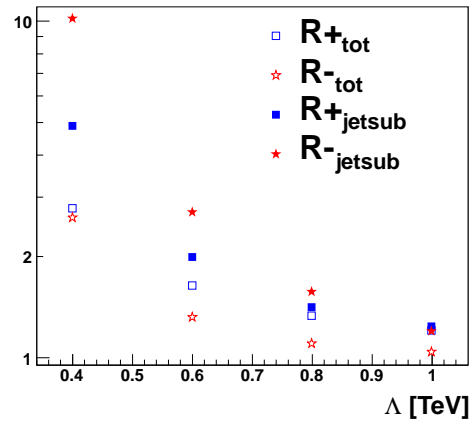}
\caption{The ratio of the cross-sections $R-$ (mixture of SM and CP
  odd ) (red stars) and $R+$ (mixture of SM and the BSM CP even term)
  (blue boxes) both before (hollow markers) and after (bold markers)
  applying selection cuts for the $WH$ channel. The scale of new physics
  $\Lambda$ also determines the strength of the contribution from new
  physics.}
\label{fig:ratio}
\end{figure}

The BSM operators of eq.~(\ref{Lagrangian}) push the $\pt$, invariant
mass $(\sqrt{\hat{s}_{HV}})$ and rapidity separation $(\Delta y_{HV})$
distributions of the $VH$ system to larger
values~\cite{Djouadi:2013yb,Ellis:2012xd,Englert:2012ct}, leading to
larger Higgs boosts and a reduced separation between the leptons
$(R_{ll})$, and $b$ jets $R_{bb}$. Consequently, the above selection
cuts enhance BSM effects. In Table~\ref{tab:cs1}, one sees the
acceptance of these operators to the selection cuts is very good:
$\sim 4$ ($\sim 6$) times the SM acceptance for the CP even (odd)
operator.

In Fig.~\ref{fig:ratio}, we consider the SM Lagrangian supplemented by
either the CP odd or even operator applied to the $WH$ channel.  We
show the ratio of the SM+BSM and SM cross-sections both for the total
cross-section
\mbox{$(R_{tot}^{\pm}=\sigma_{tot}^{SM+BSM\pm}/\sigma_{tot}^{SM})$}
and the cross-section after applying selection cuts
\mbox{$(R_{jetsub}{\pm}=\sigma_{jetsub}^{SM+BSM\pm}/\sigma_{jetsub}^{SM})$}.
As expected, the BSM contribution is larger for smaller values of the
new physics scales $\Lambda_i$.  We also see that $R_{jetsub}$
increases at a faster rate than $R_{tot}$ with decreasing values of
$\Lambda_i$. While the rates alone cannot provide information about
the $HVV$ interactions, the variation of the rates with the $\pt$ cuts
used can provide information on the presence of BSM physics. This
behaviour would also be observable in both $pp\to ZH \to l \bar{l} b
\bar b$ and $pp\to ZH \to \nu \bar{\nu} b \bar b$.

\section{Angular Observables}

\begin{figure*}
\centering
\subfloat[]{\includegraphics[width=0.35\textwidth]{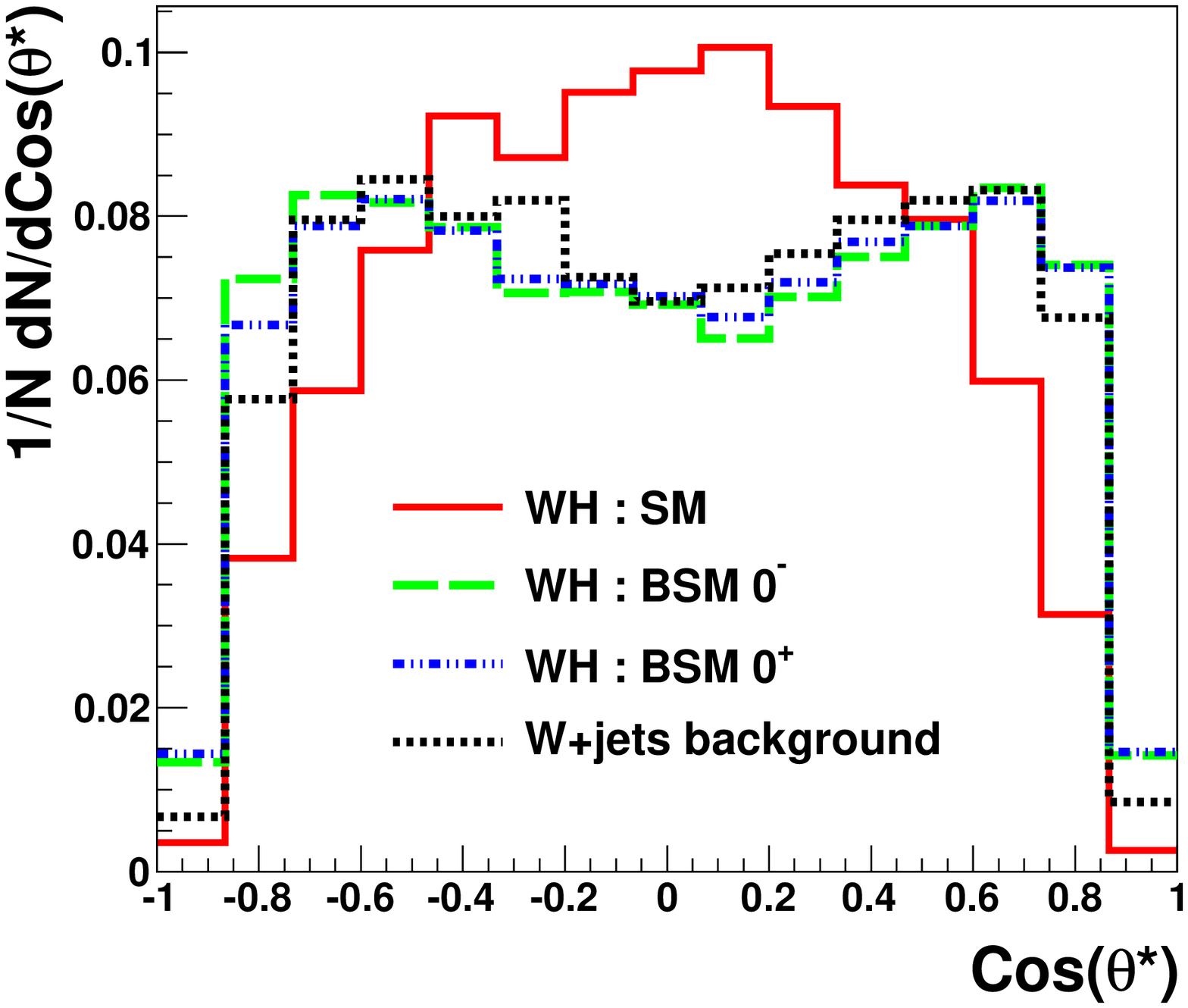}}
\subfloat[]{\includegraphics[width=0.35\textwidth]{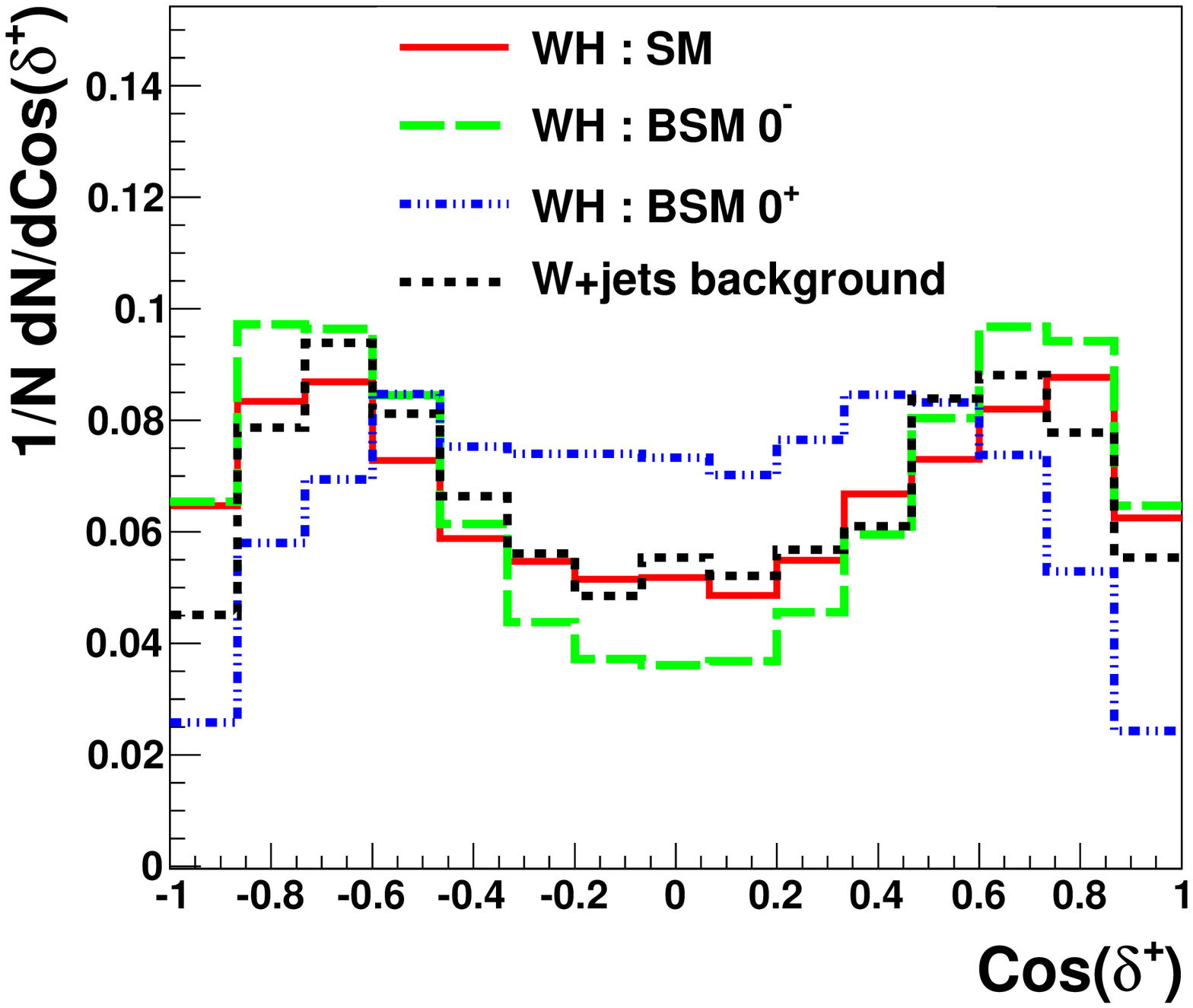}}
\subfloat[]{\includegraphics[width=0.35\textwidth]{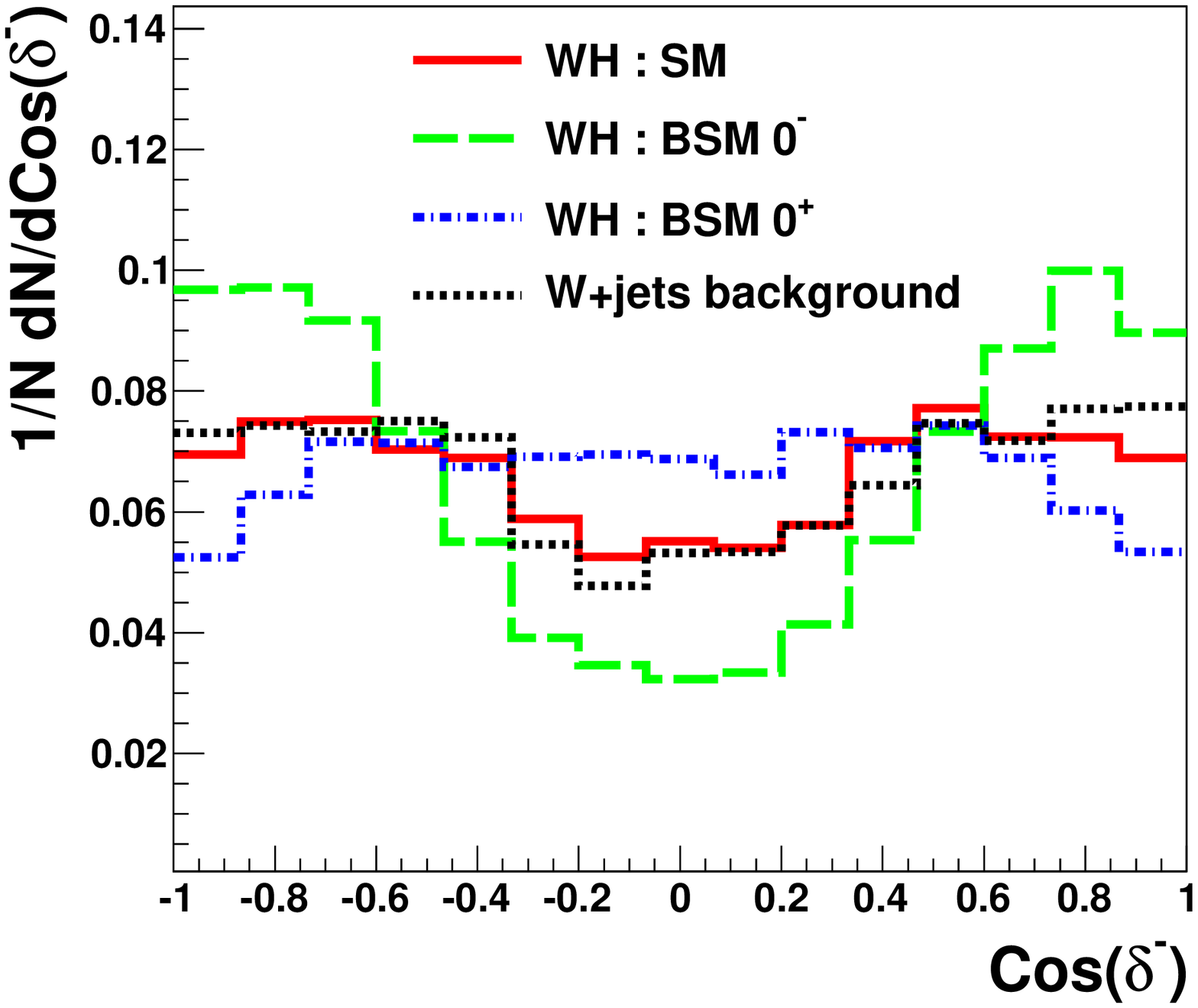}}
\caption{Distributions of the angles defined in eq.~(\ref{eqn:angles})
  for $WH$ production in the SM (solid red lines), pure BSM CP even
  (dot-dashed blue lines), pure BSM CP odd (dashed green lines) and
  for the dominant $W$ + jets background (dotted black lines). (a)
  $\cos\theta^{*}$ , (b) $\cos\delta^{+}$ (c) $\cos\delta^{-}$ , constructed with the $BT$ algorithm. }
\label{fig:angles1}
\end{figure*}

To fully distinguish CP even and odd BSM contributions, one must
construct CP-odd observables, which is difficult in
principle~\cite{Han:2009ra}. For $ZH$ production,
Ref.~\cite{Christensen:2010pf} considered two such observables,
although these are sensitive to radiation and hadronisation
corrections; Ref.~\cite{Englert:2012xt} defined observables which are
insensitive to the CP structure of BSM contributions.
Ref.~\cite{Desai:2011yj} examined $WH$ production with the decay $H \to
WW$, though the effect of the BSM CP even term was not
considered. Recently an analysis of Tevatron data used the transverse
mass to distinguish between a CP odd, spin-2 and SM
state~\cite{Johnson:2013zza}, although this does not distinguish
between the CP even and CP odd terms and has shown to be an
insensitive observable at the LHC~\cite{Ellis:2012xd}. The suggestion
in Ref.~\cite{Choi:2002jk,*Godbole:2007cn} to distinguish the CP
nature from threshold behaviour is incompatible with jet
substructure methods, which require the Higgs be highly boosted.  However,
the Lorentz structure of the BSM vertices will be reflected in the
angular distribution of the gauge boson's decay products. 
The momenta of the $V$ and Higgs bosons are reconstructed
from the leptons and jets as follows:
\begin{equation}
p_V=p_{l_1}+p_{l_2}, \quad p_H=p_{b_1}+p_{b_2} + p_{j},
\label{ZHmoms}
\end{equation}
where  $\{p_{b_i}\}$ are the momenta of the $b$ jets, $p_{j}$ is
the momentum of the light quark jet if it is reconstructed and $p_{l_1}$ and
$p_{l_2}$ are the momenta of the lepton and the anti-lepton
respectively (for $WH$, $p_{l_1}$ corresponds to the lepton momentum
and $p_{l_2}$ to the neutrino). With these momenta, we may define
\begin{eqnarray}
\cos\theta^*&=&
\frac{\vec{p}_{l_1}^{\,(V)}\cdot\vec{p}_V}{|\vec{p}_{l_1}^{\,(V)}|\,|\vec{p}_V|},
\quad 
\cos\delta^+=
\frac{\vec{p}_{l_1}^{\,(V)}\cdot\left(\vec{p}_V\times \vec{p}_H\right)}
{|\vec{p}_{l_1}^{\,(V)}|\,|\vec{p}_V\times \vec{p}_H|}, \nonumber \\
\cos\delta^- &=& 
\frac{(\vec{p}_{l_1}^{\,(H-)}\times \vec{p}_{l_2}^{\,(H-)})\cdot\vec{p}_V}
{|(\vec{p}_{l_1}^{\,(H-)}\times \vec{p}_{l_2}^{\,(H-)})||\vec{p}_V|}.
\label{eqn:angles}
\end{eqnarray}
Here $\vec{p}_{X}^{\,(Y)}$ corresponds to the three momentum of the
particle $X$ in the rest frame of the particle $Y$. If $Y$ is not
specified then the momentum is defined in the lab frame. Momenta
labels are as follows: $H$ corresponds to the Higgs boson, $H-$ stands
for the four momentum obtained when the sign of the spatial component
of the Higgs momentum is inverted $(\vec{p}_{H} \to -\vec{p}_{H})$ and
$V=W^{\pm},Z$.

Fig.~\ref{fig:angles1} shows distributions of these angles in
$WH$ production for pure SM, pure BSM and for the dominant $W$ + jets
background. The $\cos\theta^*$ distribution is the same for the
backgrounds and both BSM operators. 
The angle $\cos\theta^*$ was first defined in \cite{Miller:2001bi},
and distinguishes the SM from BSM operators, whereas $\cos\delta^{+}$ 
distinguishes the BSM CP even contribution. Finally, $\cos\delta^{-}$
distinguishes the CP odd and CP even (SM or BSM)
contributions. Results for $ZH$ production (not shown) are
qualitatively similar. Also, the
distributions for $\cos\theta^{*}$ and $\cos\delta^{+}$ are the same
if $\vec{p}_{l_1}$ is replaced by $\vec{p}_{l_2}$ in
eq.~(\ref{eqn:angles}), so one could use both leptons to improve
the statistics. 

\subsection{Asymmetries}
\label{sec:asym}

\begin{table}
\centering
\begin{tabular}{|c|c|c|c|c|}
\hline
Asymmetries          &  $ZH_{SM}$ & $ZH_{BSM}^{0-}$ & $ZH_{BSM}^{0+}$ & Z+jets \\ \hline
$A(\cos\theta^*)$    &  0.35     & -0.05         & -0.02         &  0.07  \\ \hline
$A(\cos\delta^+)$    & -0.207    & -0.262        &  0.088        & -0.188 \\\hline
$A(\cos\delta^-)$    & -0.209    & -0.435        & -0.103        & -0.321 \\ \hline
\end{tabular}
\caption{Asymmetries for the angles defined eq.~\ref{eqn:angles} in
  $ZH$ production for the SM and BSM vertices at 14 TeV LHC after
  application of all cuts.}
\label{tab:zh-asym}
\end{table}

\begin{table}
\centering
\scalebox{0.9}{
\begin{tabular}{|c|c|c|c|c|}
\hline
Asymmetries & $WH_{SM}$ & $WH_{BSM}^{0-}$ & $WH_{BSM}^{0+}$ & W+jets \\ \hline
$A(\cos\theta^*)$ & $0.396^{0.413}_{0.411}$ & $0.073^{0.082}_{0.060}$ & $0.100^{0.096}_{0.095}$ & $0.142^{0.152}_{0.132}$ \\ \hline
$A(\cos\delta^+)$ & $-0.150^{-0.204}_{-0.161}$ & $-0.284^{-0.342}_{-0.289}$ & $0.142^{0.093}_{0.141}$ & $-0.138^{-0.189}_{-0.138}$ \\\hline
$A(\cos\delta^-)$ & $-0.058^{-0.104}_{-0.059}$ & $-0.353^{-0.403}_{-0.367}$ & $0.042^{-0.003}_{0.030}$ & $-0.118^{-0.173}_{-0.135}$\\ \hline
\end{tabular}
}
\caption{Asymmetries for $WH$ production. The numbers are written as
  $BT^{MCT}_{BN}$, where $BN$, $MCT$ and $BT$ are the three
  reconstructions of the neutrino momentum. }
\label{tab:wh-asym}
\end{table}

\begin{figure}
\centering
\includegraphics[width=0.4\textwidth]{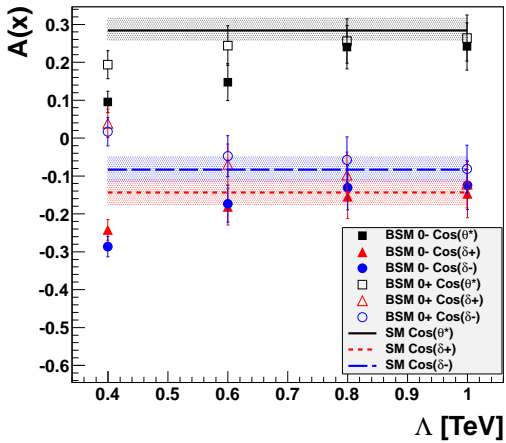}
\caption{The value of the asymmetries defined in eq.~(\ref{eqn:asym})
  for the CPC and CPV scenarios for $WH$ production, constructed with the $BT$ algorithm. 
  The strength of
  the BSM contribution is varied through the parameters
  $\Lambda_i$. The horizontal lines indicate the asymmetry in the
  SM. The contribution from the dominant $Wjj$ background is included
  in the evaluation of these asymmetries. The statistical uncertainty
  in the determination of these asymmetries for $300 \,{\rm fb}^{-1}$ of data
  for 14 TeV LHC is shown by the shaded regions for SM and by the
  error bars for BSM scenarios.}
\label{fig:asym}
\end{figure}

Motivated by Fig.~\ref{fig:angles1}, we define the asymmetry parameters
\begin{equation}
A(X)=\frac{\sigma(|X|< 0.5) - \sigma(|X|> 0.5)}{\sigma(|X|< 0.5) + \sigma(|X|> 0.5)}
\label{eqn:asym}
\end{equation}
where $X\in\{\cos\theta^*,\cos\delta^+,\cos\delta^-\}$.  The SM, pure
BSM and the dominant background are shown for $ZH$ and $WH$ production
in Tables~\ref{tab:zh-asym} and~\ref{tab:wh-asym} respectively.  We
see that $A(\cos\theta^*)$ discriminates SM from pure BSM
contributions for both $WH$ and $ZH$ production.  The other angles
discriminate the BSM CP odd and even vertices. In $WH$ production,
reconstruction ambiguities of the $W$ 
shift the absolute value of the asymmetries by roughly the same amount, cf. Table~\ref{tab:wh-asym}.
Thus differences in asymmetries are robust against this systematic uncertainty.

In Fig.~\ref{fig:asym} we show the variation of these asymmetries with
$\Lambda_i$ in $WH$ production, including the new operators with
the SM. Their mutual interference gives CP conserving (CPC) or CP violating (CPV) processes
if the CP even or odd operators are present
respectively. Our results include 
the dominant $Wjj$ background. As expected, the
asymmetries approach the SM value with increasing $\Lambda_i$. For
$\cos\theta^{*}$ the CP even and odd operators both reduce
the asymmetry, whereas for $\cos\delta{\pm}$ their
effects are of opposite sign. One can thus
effectively discriminate the BSM contributions to the
vertex. Similar results are obtained for $ZH$ production.  Fig.~\ref{fig:asym}
 includes an estimate of the statistical
uncertainty. Bounds on
the value of $\Lambda_i> 400\,{\rm GeV}$ can be easily placed and for
the CPV scenario this may be extended up to $\Lambda_2>600\,{\rm GeV}$
for $300 \,{\rm fb}^{-1}$ of LHC data. Obviously the suggested reach of $3000
\,{\rm fb}^{-1}$ of data for LHC will reduce the uncertainties and therefore
higher values of scales $\Lambda_i$ can be probed with more data. One may also combine kinematic and asymmetry information in a multivariate analysis. 

\section{Conclusions}

We examine $ZH$ and $WH$ production at the LHC, where the Higgs
decays to a $b \bar b$ pair. Combining jet substructure techniques
with vector boson polarisation (via angular distributions
of decay products), we give observables that
can distinguish between new operators coupling the Higgs to
vector bosons. Importantly (given that 
the newly discovered boson cannot be purely CP-odd), our analysis applies when
both BSM and SM operators are present, and mutually
interfere.  We show that in $VH$ production, 
sensitivity to BSM physics is enhanced through an increased acceptance
of BSM couplings to the selection cuts, and the $HWW$ and $HZZ$
couplings can be studied independently of each other. Further
investigation, including possible detector effects, is ongoing.

\section{Acknowledgements} 
We thank D. Sengupta for discussions regarding $W$ reconstruction and S. Vempati for comments on the manuscript. KM
acknowledges the financial support from CSIR India , the French CMIRA
and ENIGMASS Labex. DJM and CDW thank the Indian Institute of Science
for their hospitality while part of this work was carried out. DJM
acknowledges partial support from the Royal Society of Edinburgh, the
Indian National Science Academy and the STFC. RG wishes to thank the
Department of Science and Technology, Government of India, for support
under grant no. SR/S2/JCB-64/2007.

\bibliographystyle{apsrev4-1}
\bibliography{vhcp}

\begin{thebibliography}{34}%
\makeatletter
\providecommand \@ifxundefined [1]{%
 \@ifx{#1\undefined}
}%
\providecommand \@ifnum [1]{%
 \ifnum #1\expandafter \@firstoftwo
 \else \expandafter \@secondoftwo
 \fi
}%
\providecommand \@ifx [1]{%
 \ifx #1\expandafter \@firstoftwo
 \else \expandafter \@secondoftwo
 \fi
}%
\providecommand \natexlab [1]{#1}%
\providecommand \enquote  [1]{``#1''}%
\providecommand \bibnamefont  [1]{#1}%
\providecommand \bibfnamefont [1]{#1}%
\providecommand \citenamefont [1]{#1}%
\providecommand \href@noop [0]{\@secondoftwo}%
\providecommand \href [0]{\begingroup \@sanitize@url \@href}%
\providecommand \@href[1]{\@@startlink{#1}\@@href}%
\providecommand \@@href[1]{\endgroup#1\@@endlink}%
\providecommand \@sanitize@url [0]{\catcode `\\12\catcode `\$12\catcode
  `\&12\catcode `\#12\catcode `\^12\catcode `\_12\catcode `\%12\relax}%
\providecommand \@@startlink[1]{}%
\providecommand \@@endlink[0]{}%
\providecommand \url  [0]{\begingroup\@sanitize@url \@url }%
\providecommand \@url [1]{\endgroup\@href {#1}{\urlprefix }}%
\providecommand \urlprefix  [0]{URL }%
\providecommand \Eprint [0]{\href }%
\providecommand \doibase [0]{http://dx.doi.org/}%
\providecommand \selectlanguage [0]{\@gobble}%
\providecommand \bibinfo  [0]{\@secondoftwo}%
\providecommand \bibfield  [0]{\@secondoftwo}%
\providecommand \translation [1]{[#1]}%
\providecommand \BibitemOpen [0]{}%
\providecommand \bibitemStop [0]{}%
\providecommand \bibitemNoStop [0]{.\EOS\space}%
\providecommand \EOS [0]{\spacefactor3000\relax}%
\providecommand \BibitemShut  [1]{\csname bibitem#1\endcsname}%
\let\auto@bib@innerbib\@empty
\bibitem [{\citenamefont {Chatrchyan}\ \emph {et~al.}(2012)\citenamefont
  {Chatrchyan} \emph {et~al.}}]{Chatrchyan:2012ufa}%
  \BibitemOpen
  \bibfield  {author} {\bibinfo {author} {\bibfnamefont {S.}~\bibnamefont
  {Chatrchyan}} \emph {et~al.} (\bibinfo {collaboration} {CMS Collaboration}),\
  }\href {\doibase 10.1016/j.physletb.2012.08.021} {\bibfield  {journal}
  {\bibinfo  {journal} {Phys.Lett.}\ }\textbf {\bibinfo {volume} {B716}},\
  \bibinfo {pages} {30} (\bibinfo {year} {2012})},\ \Eprint
  {http://arxiv.org/abs/1207.7235} {arXiv:1207.7235 [hep-ex]} \BibitemShut
  {NoStop}%
\bibitem [{\citenamefont {Aad}\ \emph {et~al.}(2012)\citenamefont {Aad} \emph
  {et~al.}}]{Aad:2012tfa}%
  \BibitemOpen
  \bibfield  {author} {\bibinfo {author} {\bibfnamefont {G.}~\bibnamefont
  {Aad}} \emph {et~al.} (\bibinfo {collaboration} {ATLAS Collaboration}),\
  }\href {\doibase 10.1016/j.physletb.2012.08.020} {\bibfield  {journal}
  {\bibinfo  {journal} {Phys.Lett.}\ }\textbf {\bibinfo {volume} {B716}},\
  \bibinfo {pages} {1} (\bibinfo {year} {2012})},\ \Eprint
  {http://arxiv.org/abs/1207.7214} {arXiv:1207.7214 [hep-ex]} \BibitemShut
  {NoStop}%
\bibitem [{\citenamefont {Butterworth}\ \emph {et~al.}(2008)\citenamefont
  {Butterworth}, \citenamefont {Davison}, \citenamefont {Rubin},\ and\
  \citenamefont {Salam}}]{Butterworth:2008iy}%
  \BibitemOpen
  \bibfield  {author} {\bibinfo {author} {\bibfnamefont {J.~M.}\ \bibnamefont
  {Butterworth}}, \bibinfo {author} {\bibfnamefont {A.~R.}\ \bibnamefont
  {Davison}}, \bibinfo {author} {\bibfnamefont {M.}~\bibnamefont {Rubin}}, \
  and\ \bibinfo {author} {\bibfnamefont {G.~P.}\ \bibnamefont {Salam}},\ }\href
  {\doibase 10.1103/PhysRevLett.100.242001} {\bibfield  {journal} {\bibinfo
  {journal} {Phys.Rev.Lett.}\ }\textbf {\bibinfo {volume} {100}},\ \bibinfo
  {pages} {242001} (\bibinfo {year} {2008})},\ \Eprint
  {http://arxiv.org/abs/0802.2470} {arXiv:0802.2470 [hep-ph]} \BibitemShut
  {NoStop}%
\bibitem [{\citenamefont {Choi}\ \emph {et~al.}(2003)\citenamefont {Choi},
  \citenamefont {Miller}, \citenamefont {Muhlleitner},\ and\ \citenamefont
  {Zerwas}}]{Choi:2002jk}%
  \BibitemOpen
  \bibfield  {author} {\bibinfo {author} {\bibfnamefont {S.}~\bibnamefont
  {Choi}}, \bibinfo {author} {\bibfnamefont {D.}~\bibnamefont {Miller}},
  \bibinfo {author} {\bibfnamefont {M.}~\bibnamefont {Muhlleitner}}, \ and\
  \bibinfo {author} {\bibfnamefont {P.}~\bibnamefont {Zerwas}},\ }\href
  {\doibase 10.1016/S0370-2693(02)03191-X} {\bibfield  {journal} {\bibinfo
  {journal} {Phys.Lett.}\ }\textbf {\bibinfo {volume} {B553}},\ \bibinfo
  {pages} {61} (\bibinfo {year} {2003})},\ \Eprint
  {http://arxiv.org/abs/hep-ph/0210077} {arXiv:hep-ph/0210077 [hep-ph]}
  \BibitemShut {NoStop}%
\bibitem [{\citenamefont {Godbole}\ \emph {et~al.}(2007)\citenamefont
  {Godbole}, \citenamefont {Miller},\ and\ \citenamefont
  {Muhlleitner}}]{Godbole:2007cn}%
  \BibitemOpen
  \bibfield  {author} {\bibinfo {author} {\bibfnamefont {R.~M.}\ \bibnamefont
  {Godbole}}, \bibinfo {author} {\bibfnamefont {D.}~\bibnamefont {Miller}}, \
  and\ \bibinfo {author} {\bibfnamefont {M.~M.}\ \bibnamefont {Muhlleitner}},\
  }\href {\doibase 10.1088/1126-6708/2007/12/031} {\bibfield  {journal}
  {\bibinfo  {journal} {JHEP}\ }\textbf {\bibinfo {volume} {0712}},\ \bibinfo
  {pages} {031} (\bibinfo {year} {2007})},\ \Eprint
  {http://arxiv.org/abs/0708.0458} {arXiv:0708.0458 [hep-ph]} \BibitemShut
  {NoStop}%
\bibitem [{\citenamefont {Gao}\ \emph {et~al.}(2010)\citenamefont {Gao},
  \citenamefont {Gritsan}, \citenamefont {Guo}, \citenamefont {Melnikov},
  \citenamefont {Schulze} \emph {et~al.}}]{Gao:2010qx}%
  \BibitemOpen
  \bibfield  {author} {\bibinfo {author} {\bibfnamefont {Y.}~\bibnamefont
  {Gao}}, \bibinfo {author} {\bibfnamefont {A.~V.}\ \bibnamefont {Gritsan}},
  \bibinfo {author} {\bibfnamefont {Z.}~\bibnamefont {Guo}}, \bibinfo {author}
  {\bibfnamefont {K.}~\bibnamefont {Melnikov}}, \bibinfo {author}
  {\bibfnamefont {M.}~\bibnamefont {Schulze}},  \emph {et~al.},\ }\href
  {\doibase 10.1103/PhysRevD.81.075022} {\bibfield  {journal} {\bibinfo
  {journal} {Phys.Rev.}\ }\textbf {\bibinfo {volume} {D81}},\ \bibinfo {pages}
  {075022} (\bibinfo {year} {2010})},\ \Eprint {http://arxiv.org/abs/1001.3396}
  {arXiv:1001.3396 [hep-ph]} \BibitemShut {NoStop}%
\bibitem [{\citenamefont {De~Rujula}\ \emph {et~al.}(2010)\citenamefont
  {De~Rujula}, \citenamefont {Lykken}, \citenamefont {Pierini}, \citenamefont
  {Rogan},\ and\ \citenamefont {Spiropulu}}]{DeRujula:2010ys}%
  \BibitemOpen
  \bibfield  {author} {\bibinfo {author} {\bibfnamefont {A.}~\bibnamefont
  {De~Rujula}}, \bibinfo {author} {\bibfnamefont {J.}~\bibnamefont {Lykken}},
  \bibinfo {author} {\bibfnamefont {M.}~\bibnamefont {Pierini}}, \bibinfo
  {author} {\bibfnamefont {C.}~\bibnamefont {Rogan}}, \ and\ \bibinfo {author}
  {\bibfnamefont {M.}~\bibnamefont {Spiropulu}},\ }\href {\doibase
  10.1103/PhysRevD.82.013003} {\bibfield  {journal} {\bibinfo  {journal}
  {Phys.Rev.}\ }\textbf {\bibinfo {volume} {D82}},\ \bibinfo {pages} {013003}
  (\bibinfo {year} {2010})},\ \Eprint {http://arxiv.org/abs/1001.5300}
  {arXiv:1001.5300 [hep-ph]} \BibitemShut {NoStop}%
\bibitem [{\citenamefont {Bolognesi}\ \emph {et~al.}(2012)\citenamefont
  {Bolognesi}, \citenamefont {Gao}, \citenamefont {Gritsan}, \citenamefont
  {Melnikov}, \citenamefont {Schulze} \emph {et~al.}}]{Bolognesi:2012mm}%
  \BibitemOpen
  \bibfield  {author} {\bibinfo {author} {\bibfnamefont {S.}~\bibnamefont
  {Bolognesi}}, \bibinfo {author} {\bibfnamefont {Y.}~\bibnamefont {Gao}},
  \bibinfo {author} {\bibfnamefont {A.~V.}\ \bibnamefont {Gritsan}}, \bibinfo
  {author} {\bibfnamefont {K.}~\bibnamefont {Melnikov}}, \bibinfo {author}
  {\bibfnamefont {M.}~\bibnamefont {Schulze}},  \emph {et~al.},\ }\href
  {\doibase 10.1103/PhysRevD.86.095031} {\bibfield  {journal} {\bibinfo
  {journal} {Phys.Rev.}\ }\textbf {\bibinfo {volume} {D86}},\ \bibinfo {pages}
  {095031} (\bibinfo {year} {2012})},\ \Eprint {http://arxiv.org/abs/1208.4018}
  {arXiv:1208.4018 [hep-ph]} \BibitemShut {NoStop}%
\bibitem [{\citenamefont {Stolarski}\ and\ \citenamefont
  {Vega-Morales}(2012)}]{Stolarski:2012ps}%
  \BibitemOpen
  \bibfield  {author} {\bibinfo {author} {\bibfnamefont {D.}~\bibnamefont
  {Stolarski}}\ and\ \bibinfo {author} {\bibfnamefont {R.}~\bibnamefont
  {Vega-Morales}},\ }\href {\doibase 10.1103/PhysRevD.86.117504} {\bibfield
  {journal} {\bibinfo  {journal} {Phys.Rev.}\ }\textbf {\bibinfo {volume}
  {D86}},\ \bibinfo {pages} {117504} (\bibinfo {year} {2012})},\ \Eprint
  {http://arxiv.org/abs/1208.4840} {arXiv:1208.4840 [hep-ph]} \BibitemShut
  {NoStop}%
\bibitem [{ATL(2013)}]{ATLAS-CONF-2013-013}%
  \BibitemOpen
  \href@noop {} {\emph {\bibinfo {title} {Measurements of the properties of the
  Higgs-like boson in the four lepton decay channel with the ATLAS detector
  using 25 fb$^{-1}$ of proton-proton collision data}}},\ \bibinfo {type}
  {Tech. Rep.}\ \bibinfo {number} {ATLAS-CONF-2013-013}\ (\bibinfo
  {institution} {CERN},\ \bibinfo {address} {Geneva},\ \bibinfo {year}
  {2013})\BibitemShut {NoStop}%
\bibitem [{CMS(2013)}]{CMS-PAS-HIG-13-002}%
  \BibitemOpen
  \href@noop {} {\emph {\bibinfo {title} {Properties of the Higgs-like boson in
  the decay H to ZZ to 4l in pp collisions at sqrt s =7 and 8 TeV}}},\ \bibinfo
  {type} {Tech. Rep.}\ \bibinfo {number} {CMS-PAS-HIG-13-002}\ (\bibinfo
  {institution} {CERN},\ \bibinfo {address} {Geneva},\ \bibinfo {year}
  {2013})\BibitemShut {NoStop}%
\bibitem [{\citenamefont {Chatrchyan}\ \emph {et~al.}(2013)\citenamefont
  {Chatrchyan} \emph {et~al.}}]{Chatrchyan:2012jja}%
  \BibitemOpen
  \bibfield  {author} {\bibinfo {author} {\bibfnamefont {S.}~\bibnamefont
  {Chatrchyan}} \emph {et~al.} (\bibinfo {collaboration} {CMS Collaboration}),\
  }\href@noop {} {\bibfield  {journal} {\bibinfo  {journal} {Phys. Rev. Lett.}\
  }\textbf {\bibinfo {volume} {110}},\ \bibinfo {pages} {081803} (\bibinfo
  {year} {2013})},\ \Eprint {http://arxiv.org/abs/1212.6639} {arXiv:1212.6639
  [hep-ex]} \BibitemShut {NoStop}%
\bibitem [{\citenamefont {Plehn}\ \emph {et~al.}(2002)\citenamefont {Plehn},
  \citenamefont {Rainwater},\ and\ \citenamefont {Zeppenfeld}}]{Plehn:2001nj}%
  \BibitemOpen
  \bibfield  {author} {\bibinfo {author} {\bibfnamefont {T.}~\bibnamefont
  {Plehn}}, \bibinfo {author} {\bibfnamefont {D.~L.}\ \bibnamefont
  {Rainwater}}, \ and\ \bibinfo {author} {\bibfnamefont {D.}~\bibnamefont
  {Zeppenfeld}},\ }\href {\doibase 10.1103/PhysRevLett.88.051801} {\bibfield
  {journal} {\bibinfo  {journal} {Phys.Rev.Lett.}\ }\textbf {\bibinfo {volume}
  {88}},\ \bibinfo {pages} {051801} (\bibinfo {year} {2002})},\ \Eprint
  {http://arxiv.org/abs/hep-ph/0105325} {arXiv:hep-ph/0105325 [hep-ph]}
  \BibitemShut {NoStop}%
\bibitem [{\citenamefont {Hankele}\ \emph {et~al.}(2006)\citenamefont
  {Hankele}, \citenamefont {Klamke}, \citenamefont {Zeppenfeld},\ and\
  \citenamefont {Figy}}]{Zeppen:2006}%
  \BibitemOpen
  \bibfield  {author} {\bibinfo {author} {\bibfnamefont {V.}~\bibnamefont
  {Hankele}}, \bibinfo {author} {\bibfnamefont {G.}~\bibnamefont {Klamke}},
  \bibinfo {author} {\bibfnamefont {D.}~\bibnamefont {Zeppenfeld}}, \ and\
  \bibinfo {author} {\bibfnamefont {T.}~\bibnamefont {Figy}},\ }\href {\doibase
  10.1103/PhysRevD.74.095001} {\bibfield  {journal} {\bibinfo  {journal}
  {Phys.Rev.}\ }\textbf {\bibinfo {volume} {D74}},\ \bibinfo {pages} {095001}
  (\bibinfo {year} {2006})},\ \Eprint {http://arxiv.org/abs/hep-ph/0609075}
  {arXiv:hep-ph/0609075 [hep-ph]} \BibitemShut {NoStop}%
\bibitem [{\citenamefont {Miller}\ \emph {et~al.}(2001)\citenamefont {Miller},
  \citenamefont {Choi}, \citenamefont {Eberle}, \citenamefont {Muhlleitner},\
  and\ \citenamefont {Zerwas}}]{Miller:2001bi}%
  \BibitemOpen
  \bibfield  {author} {\bibinfo {author} {\bibfnamefont {D.}~\bibnamefont
  {Miller}}, \bibinfo {author} {\bibfnamefont {S.}~\bibnamefont {Choi}},
  \bibinfo {author} {\bibfnamefont {B.}~\bibnamefont {Eberle}}, \bibinfo
  {author} {\bibfnamefont {M.}~\bibnamefont {Muhlleitner}}, \ and\ \bibinfo
  {author} {\bibfnamefont {P.}~\bibnamefont {Zerwas}},\ }\href {\doibase
  10.1016/S0370-2693(01)00317-3} {\bibfield  {journal} {\bibinfo  {journal}
  {Phys.Lett.}\ }\textbf {\bibinfo {volume} {B505}},\ \bibinfo {pages} {149}
  (\bibinfo {year} {2001})},\ \Eprint {http://arxiv.org/abs/hep-ph/0102023}
  {arXiv:hep-ph/0102023 [hep-ph]} \BibitemShut {NoStop}%
\bibitem [{\citenamefont {Han}\ and\ \citenamefont {Jiang}(2001)}]{Han:2000mi}%
  \BibitemOpen
  \bibfield  {author} {\bibinfo {author} {\bibfnamefont {T.}~\bibnamefont
  {Han}}\ and\ \bibinfo {author} {\bibfnamefont {J.}~\bibnamefont {Jiang}},\
  }\href {\doibase 10.1103/PhysRevD.63.096007} {\bibfield  {journal} {\bibinfo
  {journal} {Phys.Rev.}\ }\textbf {\bibinfo {volume} {D63}},\ \bibinfo {pages}
  {096007} (\bibinfo {year} {2001})},\ \Eprint
  {http://arxiv.org/abs/hep-ph/0011271} {arXiv:hep-ph/0011271 [hep-ph]}
  \BibitemShut {NoStop}%
\bibitem [{\citenamefont {Rindani}\ and\ \citenamefont
  {Sharma}(2009)}]{Rindani:2009pb}%
  \BibitemOpen
  \bibfield  {author} {\bibinfo {author} {\bibfnamefont {S.~D.}\ \bibnamefont
  {Rindani}}\ and\ \bibinfo {author} {\bibfnamefont {P.}~\bibnamefont
  {Sharma}},\ }\href {\doibase 10.1103/PhysRevD.79.075007} {\bibfield
  {journal} {\bibinfo  {journal} {Phys.Rev.}\ }\textbf {\bibinfo {volume}
  {D79}},\ \bibinfo {pages} {075007} (\bibinfo {year} {2009})},\ \Eprint
  {http://arxiv.org/abs/0901.2821} {arXiv:0901.2821 [hep-ph]} \BibitemShut
  {NoStop}%
\bibitem [{\citenamefont {Biswal}\ \emph {et~al.}(2009)\citenamefont {Biswal},
  \citenamefont {Choudhury}, \citenamefont {Godbole},\ and\ \citenamefont
  {Mamta}}]{Biswal:2008tg}%
  \BibitemOpen
  \bibfield  {author} {\bibinfo {author} {\bibfnamefont {S.~S.}\ \bibnamefont
  {Biswal}}, \bibinfo {author} {\bibfnamefont {D.}~\bibnamefont {Choudhury}},
  \bibinfo {author} {\bibfnamefont {R.~M.}\ \bibnamefont {Godbole}}, \ and\
  \bibinfo {author} {\bibnamefont {Mamta}},\ }\href {\doibase
  10.1103/PhysRevD.79.035012} {\bibfield  {journal} {\bibinfo  {journal}
  {Phys.Rev.}\ }\textbf {\bibinfo {volume} {D79}},\ \bibinfo {pages} {035012}
  (\bibinfo {year} {2009})},\ \Eprint {http://arxiv.org/abs/0809.0202}
  {arXiv:0809.0202 [hep-ph]} \BibitemShut {NoStop}%
\bibitem [{\citenamefont {Biswal}\ and\ \citenamefont
  {Godbole}(2009)}]{Biswal:2009ar}%
  \BibitemOpen
  \bibfield  {author} {\bibinfo {author} {\bibfnamefont {S.~S.}\ \bibnamefont
  {Biswal}}\ and\ \bibinfo {author} {\bibfnamefont {R.~M.}\ \bibnamefont
  {Godbole}},\ }\href {\doibase 10.1016/j.physletb.2009.08.014} {\bibfield
  {journal} {\bibinfo  {journal} {Phys.Lett.}\ }\textbf {\bibinfo {volume}
  {B680}},\ \bibinfo {pages} {81} (\bibinfo {year} {2009})},\ \Eprint
  {http://arxiv.org/abs/0906.5471} {arXiv:0906.5471 [hep-ph]} \BibitemShut
  {NoStop}%
\bibitem [{\citenamefont {Dutta}\ \emph {et~al.}(2008)\citenamefont {Dutta},
  \citenamefont {Hagiwara},\ and\ \citenamefont {Matsumoto}}]{Dutta:2008bh}%
  \BibitemOpen
  \bibfield  {author} {\bibinfo {author} {\bibfnamefont {S.}~\bibnamefont
  {Dutta}}, \bibinfo {author} {\bibfnamefont {K.}~\bibnamefont {Hagiwara}}, \
  and\ \bibinfo {author} {\bibfnamefont {Y.}~\bibnamefont {Matsumoto}},\ }\href
  {\doibase 10.1103/PhysRevD.78.115016} {\bibfield  {journal} {\bibinfo
  {journal} {Phys.Rev.}\ }\textbf {\bibinfo {volume} {D78}},\ \bibinfo {pages}
  {115016} (\bibinfo {year} {2008})},\ \Eprint {http://arxiv.org/abs/0808.0477}
  {arXiv:0808.0477 [hep-ph]} \BibitemShut {NoStop}%
\bibitem [{\citenamefont {Biswal}\ \emph {et~al.}(2012)\citenamefont {Biswal},
  \citenamefont {Godbole}, \citenamefont {Mellado},\ and\ \citenamefont
  {Raychaudhuri}}]{Biswal:2012mp}%
  \BibitemOpen
  \bibfield  {author} {\bibinfo {author} {\bibfnamefont {S.~S.}\ \bibnamefont
  {Biswal}}, \bibinfo {author} {\bibfnamefont {R.~M.}\ \bibnamefont {Godbole}},
  \bibinfo {author} {\bibfnamefont {B.}~\bibnamefont {Mellado}}, \ and\
  \bibinfo {author} {\bibfnamefont {S.}~\bibnamefont {Raychaudhuri}},\ }\href
  {\doibase 10.1103/PhysRevLett.109.261801} {\bibfield  {journal} {\bibinfo
  {journal} {Phys.Rev.Lett.}\ }\textbf {\bibinfo {volume} {109}},\ \bibinfo
  {pages} {261801} (\bibinfo {year} {2012})},\ \Eprint
  {http://arxiv.org/abs/1203.6285} {arXiv:1203.6285 [hep-ph]} \BibitemShut
  {NoStop}%
\bibitem [{\citenamefont {Alwall}\ \emph {et~al.}(2011)\citenamefont {Alwall},
  \citenamefont {Herquet}, \citenamefont {Maltoni}, \citenamefont {Mattelaer},\
  and\ \citenamefont {Stelzer}}]{madgraph}%
  \BibitemOpen
  \bibfield  {author} {\bibinfo {author} {\bibfnamefont {J.}~\bibnamefont
  {Alwall}}, \bibinfo {author} {\bibfnamefont {M.}~\bibnamefont {Herquet}},
  \bibinfo {author} {\bibfnamefont {F.}~\bibnamefont {Maltoni}}, \bibinfo
  {author} {\bibfnamefont {O.}~\bibnamefont {Mattelaer}}, \ and\ \bibinfo
  {author} {\bibfnamefont {T.}~\bibnamefont {Stelzer}},\ }\href {\doibase
  10.1007/JHEP06(2011)128} {\bibfield  {journal} {\bibinfo  {journal} {JHEP}\
  }\textbf {\bibinfo {volume} {1106}},\ \bibinfo {pages} {128} (\bibinfo {year}
  {2011})},\ \Eprint {http://arxiv.org/abs/1106.0522} {arXiv:1106.0522
  [hep-ph]} \BibitemShut {NoStop}%
\bibitem [{\citenamefont {Sjostrand}\ \emph {et~al.}(2006)\citenamefont
  {Sjostrand}, \citenamefont {Mrenna},\ and\ \citenamefont {Skands}}]{pythia}%
  \BibitemOpen
  \bibfield  {author} {\bibinfo {author} {\bibfnamefont {T.}~\bibnamefont
  {Sjostrand}}, \bibinfo {author} {\bibfnamefont {S.}~\bibnamefont {Mrenna}}, \
  and\ \bibinfo {author} {\bibfnamefont {P.~Z.}\ \bibnamefont {Skands}},\
  }\href {\doibase 10.1088/1126-6708/2006/05/026} {\bibfield  {journal}
  {\bibinfo  {journal} {JHEP}\ }\textbf {\bibinfo {volume} {0605}},\ \bibinfo
  {pages} {026} (\bibinfo {year} {2006})},\ \Eprint
  {http://arxiv.org/abs/hep-ph/0603175} {arXiv:hep-ph/0603175 [hep-ph]}
  \BibitemShut {NoStop}%
\bibitem [{\citenamefont {Cacciari}\ \emph {et~al.}(2012)\citenamefont
  {Cacciari}, \citenamefont {Salam},\ and\ \citenamefont {Soyez}}]{fastjet}%
  \BibitemOpen
  \bibfield  {author} {\bibinfo {author} {\bibfnamefont {M.}~\bibnamefont
  {Cacciari}}, \bibinfo {author} {\bibfnamefont {G.~P.}\ \bibnamefont {Salam}},
  \ and\ \bibinfo {author} {\bibfnamefont {G.}~\bibnamefont {Soyez}},\ }\href
  {\doibase 10.1140/epjc/s10052-012-1896-2} {\bibfield  {journal} {\bibinfo
  {journal} {Eur.Phys.J.}\ }\textbf {\bibinfo {volume} {C72}},\ \bibinfo
  {pages} {1896} (\bibinfo {year} {2012})},\ \Eprint
  {http://arxiv.org/abs/1111.6097} {arXiv:1111.6097 [hep-ph]} \BibitemShut
  {NoStop}%
\bibitem [{\citenamefont {Christensen}\ and\ \citenamefont
  {Duhr}(2009)}]{feynrules}%
  \BibitemOpen
  \bibfield  {author} {\bibinfo {author} {\bibfnamefont {N.~D.}\ \bibnamefont
  {Christensen}}\ and\ \bibinfo {author} {\bibfnamefont {C.}~\bibnamefont
  {Duhr}},\ }\href {\doibase 10.1016/j.cpc.2009.02.018} {\bibfield  {journal}
  {\bibinfo  {journal} {Comput.Phys.Commun.}\ }\textbf {\bibinfo {volume}
  {180}},\ \bibinfo {pages} {1614} (\bibinfo {year} {2009})},\ \Eprint
  {http://arxiv.org/abs/0806.4194} {arXiv:0806.4194 [hep-ph]} \BibitemShut
  {NoStop}%
\bibitem [{\citenamefont {{LHC Higgs Cross Section Working Group}}\ \emph
  {et~al.}(2011)\citenamefont {{LHC Higgs Cross Section Working Group}},
  \citenamefont {Dittmaier}, \citenamefont {Mariotti}, \citenamefont
  {Passarino},\ and\ \citenamefont {Tanaka~(Eds.)}}]{LHCH:2011ti}%
  \BibitemOpen
  \bibfield  {author} {\bibinfo {author} {\bibnamefont {{LHC Higgs Cross
  Section Working Group}}}, \bibinfo {author} {\bibfnamefont {S.}~\bibnamefont
  {Dittmaier}}, \bibinfo {author} {\bibfnamefont {C.}~\bibnamefont {Mariotti}},
  \bibinfo {author} {\bibfnamefont {G.}~\bibnamefont {Passarino}}, \ and\
  \bibinfo {author} {\bibfnamefont {R.}~\bibnamefont {Tanaka~(Eds.)}},\
  }\href@noop {} {\bibfield  {journal} {\bibinfo  {journal} {CERN-2011-002}\ }
  (\bibinfo {year} {CERN, Geneva, 2011})},\ \Eprint
  {http://arxiv.org/abs/1101.0593} {arXiv:1101.0593 [hep-ph]} \BibitemShut
  {NoStop}%
\bibitem [{\citenamefont {Djouadi}\ \emph {et~al.}(2013)\citenamefont
  {Djouadi}, \citenamefont {Godbole}, \citenamefont {Mellado},\ and\
  \citenamefont {Mohan}}]{Djouadi:2013yb}%
  \BibitemOpen
  \bibfield  {author} {\bibinfo {author} {\bibfnamefont {A.}~\bibnamefont
  {Djouadi}}, \bibinfo {author} {\bibfnamefont {R.}~\bibnamefont {Godbole}},
  \bibinfo {author} {\bibfnamefont {B.}~\bibnamefont {Mellado}}, \ and\
  \bibinfo {author} {\bibfnamefont {K.}~\bibnamefont {Mohan}},\ }\href@noop {}
  {\  (\bibinfo {year} {2013})},\ \Eprint {http://arxiv.org/abs/1301.4965}
  {arXiv:1301.4965 [hep-ph]} \BibitemShut {NoStop}%
\bibitem [{\citenamefont {Ellis}\ \emph {et~al.}(2012)\citenamefont {Ellis},
  \citenamefont {Hwang}, \citenamefont {Sanz},\ and\ \citenamefont
  {You}}]{Ellis:2012xd}%
  \BibitemOpen
  \bibfield  {author} {\bibinfo {author} {\bibfnamefont {J.}~\bibnamefont
  {Ellis}}, \bibinfo {author} {\bibfnamefont {D.~S.}\ \bibnamefont {Hwang}},
  \bibinfo {author} {\bibfnamefont {V.}~\bibnamefont {Sanz}}, \ and\ \bibinfo
  {author} {\bibfnamefont {T.}~\bibnamefont {You}},\ }\href {\doibase
  10.1007/JHEP11(2012)134} {\bibfield  {journal} {\bibinfo  {journal} {JHEP}\
  }\textbf {\bibinfo {volume} {1211}},\ \bibinfo {pages} {134} (\bibinfo {year}
  {2012})},\ \Eprint {http://arxiv.org/abs/1208.6002} {arXiv:1208.6002
  [hep-ph]} \BibitemShut {NoStop}%
\bibitem [{\citenamefont {Englert}\ \emph {et~al.}(2012)\citenamefont
  {Englert}, \citenamefont {Spannowsky},\ and\ \citenamefont
  {Takeuchi}}]{Englert:2012ct}%
  \BibitemOpen
  \bibfield  {author} {\bibinfo {author} {\bibfnamefont {C.}~\bibnamefont
  {Englert}}, \bibinfo {author} {\bibfnamefont {M.}~\bibnamefont {Spannowsky}},
  \ and\ \bibinfo {author} {\bibfnamefont {M.}~\bibnamefont {Takeuchi}},\
  }\href {\doibase 10.1007/JHEP06(2012)108} {\bibfield  {journal} {\bibinfo
  {journal} {JHEP}\ }\textbf {\bibinfo {volume} {1206}},\ \bibinfo {pages}
  {108} (\bibinfo {year} {2012})},\ \Eprint {http://arxiv.org/abs/1203.5788}
  {arXiv:1203.5788 [hep-ph]} \BibitemShut {NoStop}%
\bibitem [{\citenamefont {Han}\ and\ \citenamefont {Li}(2010)}]{Han:2009ra}%
  \BibitemOpen
  \bibfield  {author} {\bibinfo {author} {\bibfnamefont {T.}~\bibnamefont
  {Han}}\ and\ \bibinfo {author} {\bibfnamefont {Y.}~\bibnamefont {Li}},\
  }\href {\doibase 10.1016/j.physletb.2009.12.047} {\bibfield  {journal}
  {\bibinfo  {journal} {Phys.Lett.}\ }\textbf {\bibinfo {volume} {B683}},\
  \bibinfo {pages} {278} (\bibinfo {year} {2010})},\ \Eprint
  {http://arxiv.org/abs/0911.2933} {arXiv:0911.2933 [hep-ph]} \BibitemShut
  {NoStop}%
\bibitem [{\citenamefont {Christensen}\ \emph {et~al.}(2010)\citenamefont
  {Christensen}, \citenamefont {Han},\ and\ \citenamefont
  {Li}}]{Christensen:2010pf}%
  \BibitemOpen
  \bibfield  {author} {\bibinfo {author} {\bibfnamefont {N.~D.}\ \bibnamefont
  {Christensen}}, \bibinfo {author} {\bibfnamefont {T.}~\bibnamefont {Han}}, \
  and\ \bibinfo {author} {\bibfnamefont {Y.}~\bibnamefont {Li}},\ }\href
  {\doibase 10.1016/j.physletb.2010.08.008} {\bibfield  {journal} {\bibinfo
  {journal} {Phys.Lett.}\ }\textbf {\bibinfo {volume} {B693}},\ \bibinfo
  {pages} {28} (\bibinfo {year} {2010})},\ \Eprint
  {http://arxiv.org/abs/1005.5393} {arXiv:1005.5393 [hep-ph]} \BibitemShut
  {NoStop}%
\bibitem [{\citenamefont {Englert}\ \emph {et~al.}(2013)\citenamefont
  {Englert}, \citenamefont {Goncalves-Netto}, \citenamefont {Mawatari},\ and\
  \citenamefont {Plehn}}]{Englert:2012xt}%
  \BibitemOpen
  \bibfield  {author} {\bibinfo {author} {\bibfnamefont {C.}~\bibnamefont
  {Englert}}, \bibinfo {author} {\bibfnamefont {D.}~\bibnamefont
  {Goncalves-Netto}}, \bibinfo {author} {\bibfnamefont {K.}~\bibnamefont
  {Mawatari}}, \ and\ \bibinfo {author} {\bibfnamefont {T.}~\bibnamefont
  {Plehn}},\ }\href {\doibase 10.1007/JHEP01(2013)148} {\bibfield  {journal}
  {\bibinfo  {journal} {JHEP}\ }\textbf {\bibinfo {volume} {1301}},\ \bibinfo
  {pages} {148} (\bibinfo {year} {2013})},\ \Eprint
  {http://arxiv.org/abs/1212.0843} {arXiv:1212.0843 [hep-ph]} \BibitemShut
  {NoStop}%
\bibitem [{\citenamefont {Desai}\ \emph {et~al.}(2011)\citenamefont {Desai},
  \citenamefont {Ghosh},\ and\ \citenamefont {Mukhopadhyaya}}]{Desai:2011yj}%
  \BibitemOpen
  \bibfield  {author} {\bibinfo {author} {\bibfnamefont {N.}~\bibnamefont
  {Desai}}, \bibinfo {author} {\bibfnamefont {D.~K.}\ \bibnamefont {Ghosh}}, \
  and\ \bibinfo {author} {\bibfnamefont {B.}~\bibnamefont {Mukhopadhyaya}},\
  }\href {\doibase 10.1103/PhysRevD.83.113004} {\bibfield  {journal} {\bibinfo
  {journal} {Phys.Rev.}\ }\textbf {\bibinfo {volume} {D83}},\ \bibinfo {pages}
  {113004} (\bibinfo {year} {2011})},\ \Eprint {http://arxiv.org/abs/1104.3327}
  {arXiv:1104.3327 [hep-ph]} \BibitemShut {NoStop}%
\bibitem [{\citenamefont {Johnson}(2013)}]{Johnson:2013zza}%
  \BibitemOpen
  \bibfield  {author} {\bibinfo {author} {\bibfnamefont {E.}~\bibnamefont
  {Johnson}},\ }\href@noop {} {\  (\bibinfo {year} {2013})},\ \Eprint
  {http://arxiv.org/abs/1305.3675} {arXiv:1305.3675 [hep-ex]} \BibitemShut
  {NoStop}%
\end{thebibliography}%

\end{document}